# Creep in Rotating Composite Disk having Variable Thickness Subjected to Thermal as well as Particle Gradient

Harjot Kaur[1,a)], Nishi Gupta[1,b)], Anjali Jain[1,c)]

[1]*Department of Mathematics (UIS), Chandigarh University, Gharuan, Mohali, Punjab 140413, India*

[a)] Corresponding author: harjotmangat030@gmail.com
[b)] nishi.gupta.phd81@gmail.com
[c)] janjali580@gmail.com

**Abstract.** The motivation behind this paper is to present the effect of thermal and particle gradient in rotating composite disk with variable thickness using Sherby's law. The values of tangential, radial stresses and strain rates are calculated at different radius using mathematical modeling. It has been observed that with increase in the variable thickness and particle gradient the stresses and strain rates decrease and determined that the creep deformation decreases with fluctuating thickness.

## INTRODUCTION

Due to high performance light weight design the rotating disk are used in many engineering applications to achieve higher operations speed. Rotating disk are numerator applications in aerospace industry for example gas turbine. The rotating disk in gas turbine is simultaneously subjected to mechanical and thermal loads. Mostly, composite disk is categorized into two types based on the geometry that are rim disks and solid disks [1]. Rotating disk in engineering has lot of attention because they are used in wide range of instruments can be either mechanical or electronic instruments [2]. The creep deformation experiences by disk material because of rigorous thermal and mechanical loading which created big effect to disk performance [3]. Creep deformations can be reduced by varying thickness, properties of disk by using functionally graded material and by doing this we are able to reduce the cost and increases efficiency of disk .To operate in high temperature these material are designed and developed[4][5].

Grag et al. [6] studied secondary state deformation during a rotating FGM disk having linearly varying thickness in existence of the thermal gradient in which disk made from composite contained silicon carbide particle in a matrix of pure aluminum. They concluded that tangential stresses increased near inner radii but decreased towards outer radii but radial stresses in rotating disk increased with increased in thermal gradient. On other side the strain rates decrease significantly.

Gupta and Singh [7] studied stable state the creep response during a isotropic FGM/ non-FGM rotating disk with uniform temperature and varying thickness by using Sherby's law. They compared isotropic non-FGM and FGM constant thickness with isotropic varying thickness disk with similar average particle content. They concluded the result that distribution of stresses and strain rates will become more constant within the isotropic FGM hyperbolic thickness disk.

Deepak et al. [8] studied creep behavior of the rotating disk made up of FGM having linearly varying thickness. They find the strain and stress rates for particle gradient with constant temperature. They observed that with increases particle gradient the radial stress increases and tangential stress increases from inner radii but decrease at outer radii. They observed that stress and strain rates in particle gradient as compare to constant distribution of reinforcement.

Khanna et al. [9]studied the stable state creep in rotating disk having different thickness profile and reinforcement gradients. They concluded that there's lesser distortion when higher thickness and higher reinforcement gradient in composite disk.

Dharmpal et al. [3]studied the stable state deformation in variable thickness rotating disk made from FGM Al-SiC$_P$ for three different profiles constant thickness, hyperbolic varying and linearly varying thickness. They concluded the result that stresses and strain rate decreases in hyperbolic thickness profile as compare to other thickness profile. They observed that linear and hyperbolic thickness profile a little chance of damage compares with constant thickness.

Garg et al.[10] studied the secondary state creep behavior of a FGM rotating disk in linear varying thickness with thermal gradient by using the von Mises yield criteria along the radial distance at the different value of thermal gradient. They concluded that increases the thermal gradient the value radial stresses increases with increases entire radius but tangential stresses increase near inner radius and begin decreases against outer radius but strain rates decrease with increases thermal gradient significantly therefore reduces the chance of distortion in FGM disk.

In current study, secondary state creep of rotating disk is assumed with varying thickness at angular speed $\omega = 16000$. Four different cases for varying thickness and different volume content are discussed. In these four cases two disks are compared. In the first disk, the disk is operating under thermal gradient 150K having internal temperature 620K and external 470K, in the second disk the disk is operating under thermal gradient 120K having internal temperature 620K and external temperature 500K.

- Case I: Volume content 5% where $V_{max} = 15\%$ and $V_{min} = 10\%$ with thickness $h_a = 2.50mm$ and $h_b = 0.50mm$.
- Case II: Volume content 10% where $V_{max} = 20\%$ and $V_{min} = 10\%$ with thickness $h_a = 2.50mm$ and $h_b = 0.50mm$.
- Case III: Volume content 5% where $V_{max} = 15\%$ and $V_{min} = 10\%$ with thickness $h_a = 44.22mm$ and $h_b = 12.97mm$.
- Case IV: Volume content 10% where $V_{max} = 20\%$ and $V_{min} = 10\%$ with thickness $h_a = 44.22mm$ and $h_b = 12.97mm$.

## MATHEMATICAL MODELING

### LINEARLY VARYING THICKNESS

Let h is the thickness of composite disk having linearly varying thickness is assumed in the form
$$h(r) = h_b + 2c(b - r) \quad (1)$$
where $c = \frac{(h_a - h_b)}{2(b-a)}$ is the slope of disk, $h_a$ and $h_b$ be the thickness at inner and outer radii respectively.

Let $I$ moment of inertia of disk at inner radii and $I_0$ the moment of inertia of disk at outer radii r and b, respectively, $A$ at inner radius a and $A_0$ outer radius b the area of the cross section of disk respectively,
$$I = \int_a^r hr^2 dr, \quad I_0 = \int_a^b hr^2 dr \quad (2)$$
$$\text{and} \quad A = \int_a^r h\, dr, \quad A_0 = \int_a^b h\, dr \quad (3)$$

The average tangential stress in disk is given below,

$$\sigma_{\theta avg} = \frac{1}{A_0} \int_a^b h\sigma_\theta dr \qquad (4)$$

## REINFORCEMENT DISTRIBUTION IN THE DISK

The composition variation in term of the volume percent of SiC, along radial distance, the V(r) is given below
$$V(r) = C - Dr, \ a < r < b \qquad (5)$$

Where, $C = V_{max} + aD$ \qquad (6)

$$D = \frac{V_{max} - V_{min}}{b-a} \qquad (7)$$

Where $V_{max}$ maximum particle content and $V_{min}$ is minimum particle contents, at inner and outer radius of the disk respectively.

The density variation $\rho(r)$ in composite disk according to law of mixture may be expressed as
$$\rho(r) = \rho_m + (\rho_d - \rho_m)\frac{V(r)}{100} \qquad (8)$$

Where $\rho_m = 3210 kg/m^3$ density of the aluminum is based matrix and $\rho_d = 2698.4 kg/m^3$ is the density of dispersed silicon carbide particle respectively,

Put value of $V(r)$ from eqⁿ (5) in eqⁿ (8) we get,
$$\rho(r) = \rho_m + (\rho_d - \rho_m)\left(\frac{C - Dr}{100}\right) = A_\rho - B_\rho \qquad (9)$$

Where, $A_\rho = \rho_m + \left(\frac{(\rho_d - \rho_m)C}{100}\right) \qquad (10)$

$$B_\rho = \frac{D(\rho_d - \rho_m)}{100} \qquad (11)$$

## THERMAL GRADIENT

Thermal gradient is considering varied from inner to outer radius is given below,
$$T(r) = E - Fr, a < r < b \qquad (12)$$
$$E = \frac{bT_{max} - aT_{min}}{b-a} \qquad (13)$$
$$F = \frac{T_{max} - T_{min}}{b-a} \qquad (14)$$

## CREEP LAW

The steady state creep response composite of varying composition have been describe by Sherby's law
$$\dot{\epsilon} = [M(\bar{\sigma} - \sigma_0)]^n \qquad (15)$$
Where n=5, $\bar{\sigma}$ is effective stress, $\sigma_0$ threshold stress and $\dot{\epsilon}$ is effective strain rates
Creep specification is given by:
$$M = \frac{1}{E}\left[\frac{AD_L\lambda^3}{|\vec{b}_r^5|}\right]^{1/n} \qquad (16)$$

Where material creep constant is $M$, $A$ is constant, $E$ is young's modulus, $|\vec{b}_r|$ magnitude of Burger's vector, $D_L$ is lattice diffusivity. Now, $M$ and $\sigma_0$ the values of creep specifications has been obtained from the creep results report for $Al - SiC_p$ composite [11], that are fixed by the following regression equation which is function of the particle size is P=1.7$\mu$m, T is the temperature and V is the volume [12].
$$\ln(M(r)) = (-34.91) + 0.2112\ln P + 4.89\ln T(r) - 0.5891\ln V(r) \qquad (17)$$
$$\sigma_0(r) = -0.02050P + 0.0378T(r) + 1.033V(r) - 4.9695 \qquad (18)$$

# MATHEMATICAL FORMULATION

Consider a disk of $Al - SiC_p$ having internal radius a, external radius b revolving at angular speed $\omega = 16000$. For an isotropic rotating disk the generalized constitutive equations are:

$$\dot{\epsilon}_r(r) = \frac{\dot{\bar{\epsilon}}}{2\bar{\sigma}}[2\sigma_r(r) - (\sigma_\theta(r) + \sigma_z(r))] \qquad (19)$$

$$\dot{\epsilon}_\theta(r) = \frac{\dot{\bar{\epsilon}}}{2\bar{\sigma}}[2\sigma_\theta(r) - (\sigma_z(r) + \sigma_r(r))] \qquad (20)$$

$$\dot{\epsilon}_z(r) = \frac{\dot{\bar{\epsilon}}}{2\bar{\sigma}}[2\sigma_z(r) - (\sigma_r(r) + \sigma_\theta(r))] \qquad (21)$$

Where strain rates are $\dot{\epsilon}_r, \dot{\epsilon}_\theta, \dot{\epsilon}_z$ and strain rates $\sigma_r, \sigma_\theta, \sigma_z$ corresponding in direction $r, \theta, z$ which indicates by subscripts. For the biaxial state of stress ($\sigma_z = 0$). The effective stress $\bar{\sigma}$ is given below where assumed that effective stress depend on von Mises criterion are given below for biaxial state of stress

$$\bar{\sigma} = \frac{1}{\sqrt{2}}[\sigma_r^2(r) + \sigma_\theta^2(r) + (\sigma_r(r) - \sigma_\theta(r))^2]^{\frac{1}{2}} \qquad (22)$$

Using equation (21) and (16) in equation (19) & (20) we get,

$$\dot{\epsilon}_r(r) = \frac{[M(r)(\bar{\sigma} - \sigma_0(r))]^n (2x(r) - 1)}{2[x^2(r) - x(r) + 1]^{\frac{1}{2}}} \qquad (23)$$

$$\dot{\epsilon}_\theta(r) = \frac{[M(r)(\bar{\sigma} - \sigma_0(r))]^n (2 - x(r))}{2[x^2(r) - x(r) + 1]^{\frac{1}{2}}} \qquad (24)$$

$$\dot{\epsilon}_z(r) = -(\dot{\epsilon}_r(r) + \dot{\epsilon}_\theta(r)) \qquad (25)$$

$$\text{Where}, x(r) = \frac{\sigma_r(r)}{\sigma_\theta(r)} \qquad (26)$$

At any radius r ratio of radial stress and tangential stress

Solving eq$^n$ (21) & eq$^n$ (22) we get,

$$\sigma_\theta(r) = \frac{\dot{u}_a^{\frac{1}{n}}}{M(r)} \varphi_1(r) + \varphi_2(r) \qquad (27)$$

Where,

$$\varphi_1(r) = \frac{\varphi(r)}{[x^2(r) - x(r) + 1]^{1/2}} \qquad (28)$$

$$\varphi_2(r) = \frac{\sigma_0(r)}{[x^2(r) - x(r) + 1]^{1/2}} \qquad (29)$$

$$\varphi(r) = \left[\frac{2(x^2(r) - x(r) + 1)^{1/2}}{r(2 - x(r))} \exp\left(\int_a^r \frac{\emptyset(r)}{r} dr\right)\right]^{1/n} \qquad (30)$$

Where,

$$\emptyset(r) = \frac{2x(r) - 1}{2 - x(r)} \qquad (31)$$

The forces equilibrium equation maybe written as,

$$\frac{d}{dr}[h(r)r\sigma_r] - h(r)\sigma_\theta + \rho(r)\omega^2 r^2 h(r) = 0 \qquad (32)$$

Integrating equation (32) taking limit between a to b using the boundary conditions $\sigma_r = 0$ at $r = a$ and $\sigma_r = 0$ at $r = b$ we get,

$$\int_a^b h(r)\sigma_\theta dr = \omega^2 \left[A_\rho I_0 - B_\rho \left(\frac{(h_b+2cb)(b^4-a^4)}{4} - \frac{2c(b^5-a^5)}{5}\right)\right] \tag{33}$$

Multiplying by h(r)dr the equation (27) then integrating resultant eq$^n$ taking limits between a to b, we find

$$\dot{u}_a^{\frac{1}{n}} = \frac{1}{\int_a^b \frac{h(r)\varphi_1(r)dr}{M(r)}} \left[A_0 \sigma_{\theta avg} - \int_a^b h(r)\varphi_2(r)dr\right] \tag{34}$$

Dividing the equation (33) by $A_0$ and noting equation (4) we obtain,

$$\sigma_{\theta avg} = \frac{1}{A_0}\int_a^b h\sigma_\theta dr = \frac{\omega^2}{A_0}\left[A_\rho I_0 - B_\rho \left(\frac{(h_b+2cb)(b^4-a^4)}{4} - \frac{2c(b^5-a^5)}{5}\right)\right] \tag{35}$$

After knowing $\sigma_{\theta avg}$, the tangential stresses ($\sigma_\theta$) is obtained from substitute the value of $\dot{u}_a^{\frac{1}{n}}$ from equation (34) into equation (27) as follows,

$$\sigma_\theta(r) = \frac{\varphi_1(r)\left[A_0 \sigma_{\theta avg} - \int_a^b h(r)\varphi_2(r)dr\right]}{M(r)\int_a^b \frac{h(r)\varphi_1(r)dr}{M(r)}} + \varphi_2(r) \tag{36}$$

Integration equation (32) between the limits a to r, radial stresses is,

$$\sigma_r(r) = \frac{1}{rh(r)}\left[\int_a^r h(r)\sigma_\theta dr - \omega^2 A_\rho I - \omega^2 B_\rho \left(\frac{(h_b+2cb)(r^4-a^4)}{4} - \frac{2c(r^5-a^5)}{5}\right)\right] \tag{37}$$

## GRAPHICAL REPRESENTATION AND DISCUSSION

The effect of varying thickness with thermal as well as particle gradient as shown in figures below:

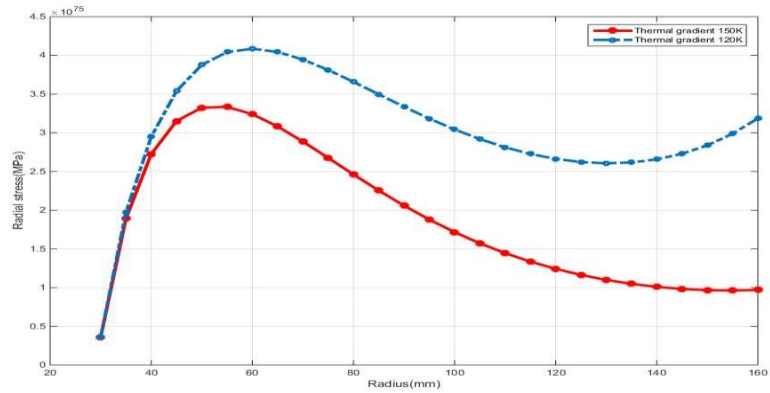

**FIGURE 1**: The variations of radial stresses along radius in rotating composite disk

The graph drawn between the radial stresses along radius with thermal gradient 150K and 120K having volume content 5% also with thickness $h_a = 2.50$ and $h_b = 0.50$. The **FIGURE 1** shows that with varying thickness the radial stresses decrease with increases thermal gradient.

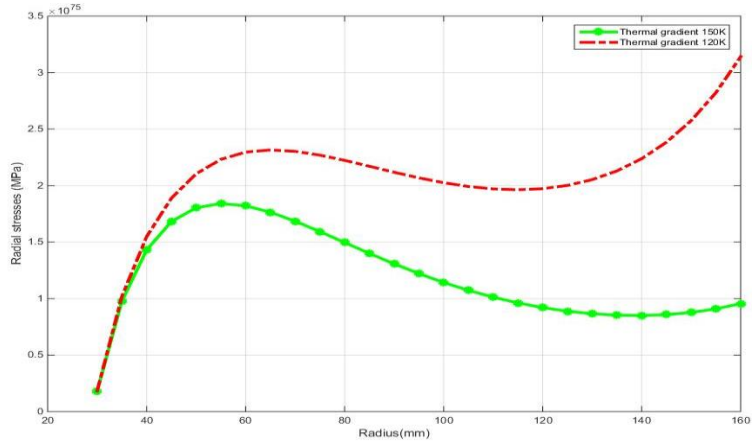

**FIGURE 2**: The variations of the radial stresses along the radius in rotating composite disk

The graph drawn between the radial stresses along radius with thermal gradient 150K and 120K having volume content 10% also with thickness $h_a = 2.50$ and $h_b = 0.50$. The **FIGURE 2** shows that with varying thickness the radial stresses decrease with increases thermal gradient. As compare to **FIGURE1** we observed in **FIGURE 2** the radial stresses value decreases with increasing volume content.

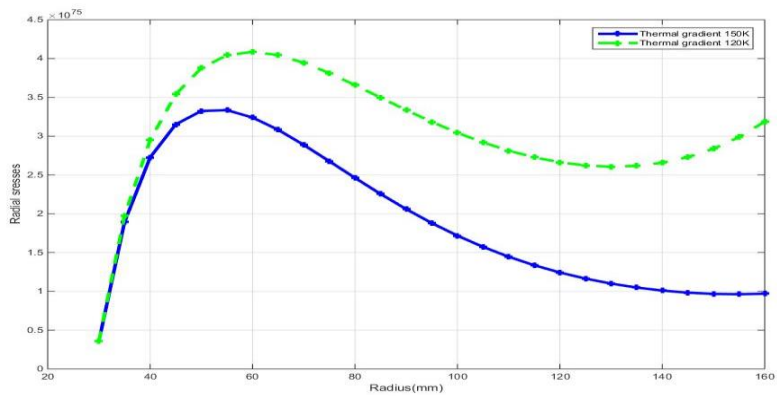

**FIGURE 3**: The variations of the radial stresses along radius in rotating composite disk

The graph drawn between the radial stresses along radius with thermal gradient 150K and 120K having volume content 5% also with thickness $h_a = 44.22$ and $h_b = 12.97$. The **FIGURE 3** shows that with varying thickness the radial stresses decreases with increases thermal gradient.

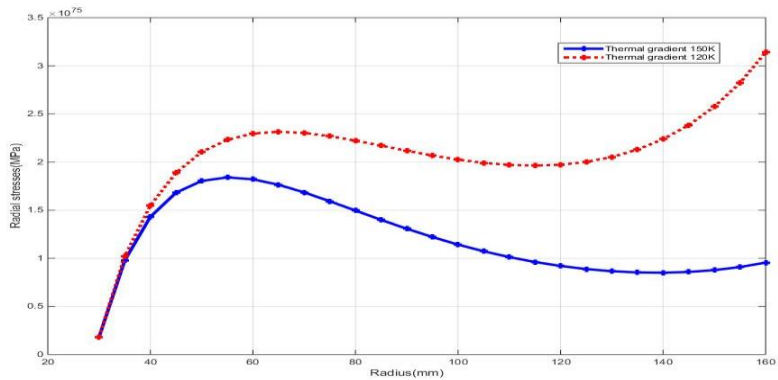

**FIGURE 4**: The variations of the radial stresses along the radius in rotating composite disk

The graph drawn between the radial stresses along radius with thermal gradient 150K and 120K having volume content 10% also with thickness $h_a = 44.22$ and $h_b = 12.97$. The **FIGURE 4** shows that with varying thickness the radial stresses decrease with increases thermal gradient. As compare to **FIGURE 3** we observed in figure4 the radial stresses value decreases.

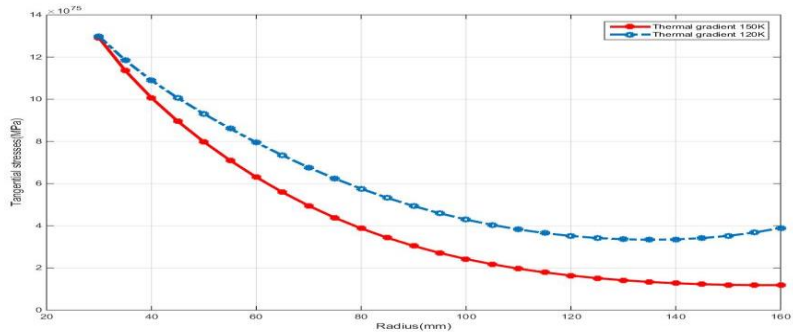

**FIGURE 5**: The variations of the tangential stresses along radius in rotating composite disk

The graph drawn between the tangential stresses along radius with thermal gradient 150K and 120K having volume content 5% also with thickness $h_a = 2.50$ and $h_b = 0.50$. The **FIGURE 5** shows that with varying thickness the tangential stresses decreases with increases thermal gradient.

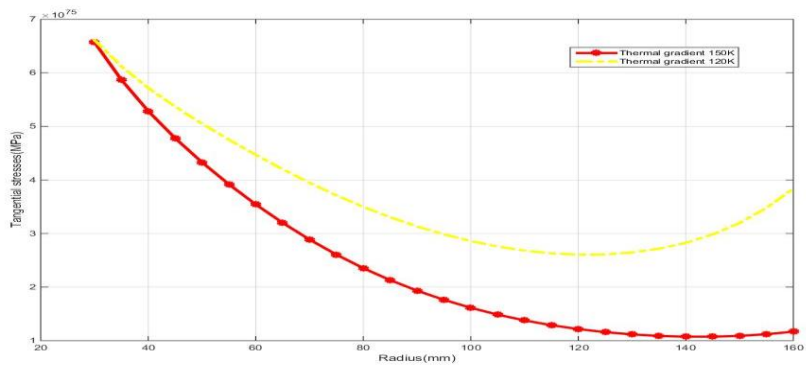

**FIGURE 6**: The variations of the tangential stresses along radius in rotating composite disk

The graph drawn between the tangential stresses along radius with thermal gradient 150K and 120K having volume content 10% also with thickness $h_a = 2.50$ and $h_b = 0.50$. The **FIGURE 6** shows that with varying thickness the tangential stresses decrease with increases thermal gradient. As compare to figure5 we observed in **FIGURE 6** the tangential stresses value decreases.

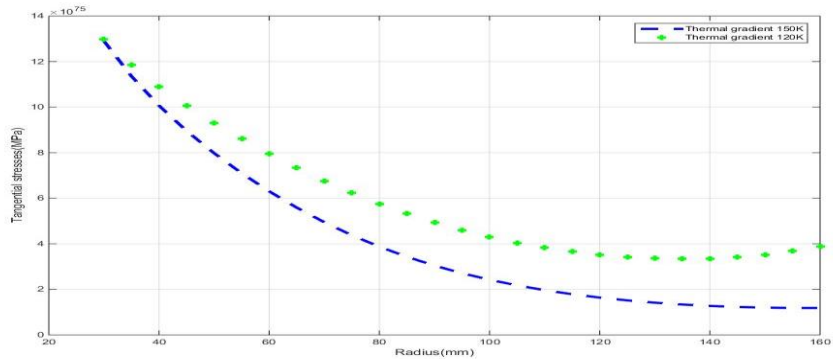

**FIGURE 7**: The variations of the tangential stresses along radius in rotating composite disk

The graph drawn between the tangential stresses along radius with thermal gradient 150K and 120K having volume content 5% also with thickness $h_a = 44.22$ and $h_b = 12.97$. The **FIGURE 7** shows that with varying thickness the tangential stresses decrease with increases thermal gradient.

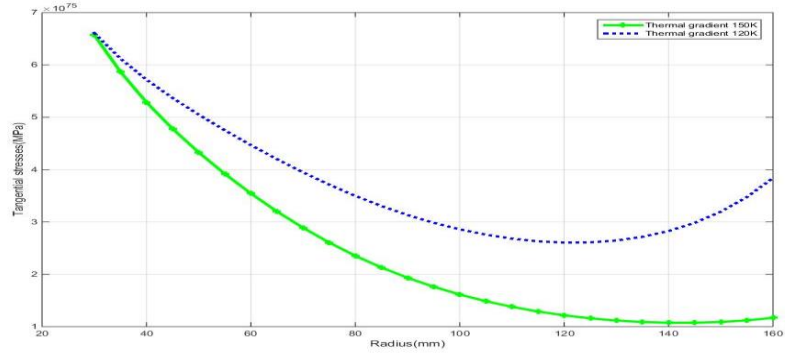

**FIGURE 8**: The variations of the tangential stresses along radius in rotating composite disk

The graph drawn between the tangential stresses along radius with thermal gradient 150K and 120K having volume content 10% also with thickness $h_a = 44.22$ and $h_b = 12.97$. The **FIGURE 8** shows that with varying thickness the tangential stresses decrease with increases thermal gradient.

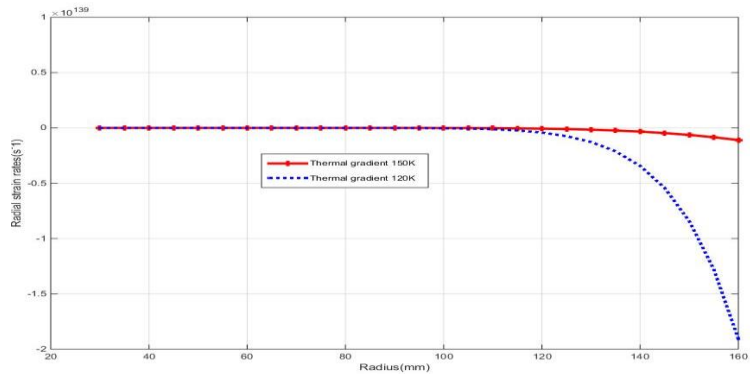

**FIGURE 9**: The variations of the radial strain rates along radius in rotating composite disk

The graph drawn between the radial strain rates along radius with thermal gradient 150K and 120K having volume content 5% also with thickness $h_a = 2.50$ and $h_b = 0.50$. The **FIGURE 9** shows that with varying thickness the radial strain rates increases with increases thermal gradient.

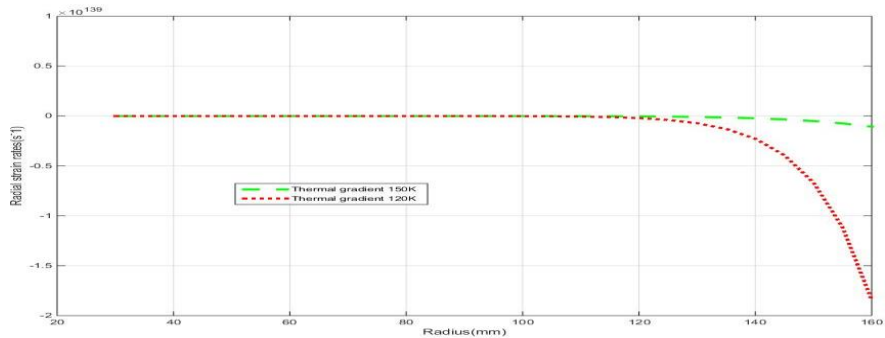

**FIGURE 10**: The variations of the radial strain rates along radius in rotating composite disk

The graph drawn between the radial strain rates along radius with thermal gradient 150K and 120K having volume content 10% also with thickness $h_a = 2.50$ and $h_b = 0.50$. The **FIGURE 10** shows that with varying thickness the radial strain rates increases with increases thermal gradient. We observed that with increases volume content with varying thickness the value of radial strain rates increases as shown in **FIGURE 10** as compared to **FIGURE 9.**

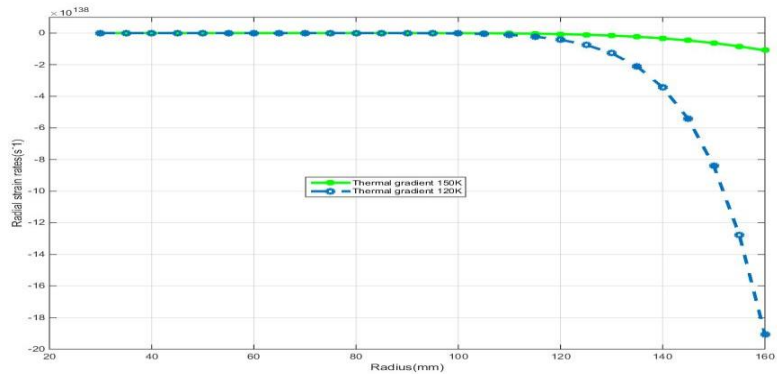

**FIGURE 11**: The variations of the radial strain rates along radius in rotating composite disk

The graph drawn between the radial strain rates along radius with thermal gradient 150K and 120K having volume content 5% also with thickness $h_a = 44.22$ and $h_b = 12.97$. The **FIGURE 11** shows that with varying thickness the radial strain rates increases with increases thermal gradient.

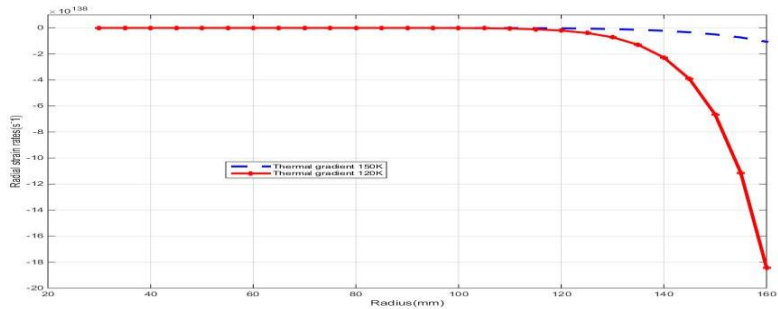

**FIGURE 12**: The variations of the radial strain rates along radius in rotating composite disk

The graph drawn between the radial strain rates along radius with thermal gradient 150K and 120K having volume content 10% also with thickness $h_a = 44.22$ and $h_b = 12.97$. The **FIGURE 12** shows that with varying thickness the radial strain rates increases with increases thermal gradient. We observed that the radial strain rates increase with increases volume content in **FIGURE 12** as compared to **FIGURE 11**.

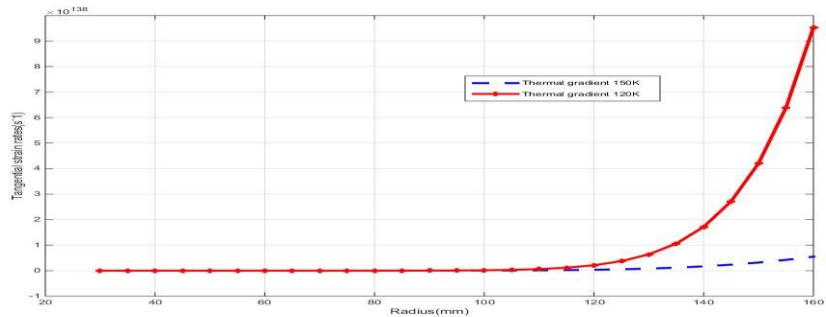

**FIGURE 13:** The variations of the tangential strain rates along radius in rotating composite disk

The graph drawn between the tangential strain rates along radius with thermal gradient 150K and 120K having volume content 5% also with thickness $h_a = 2.50$ and $h_b = 0.50$. The **FIGURE 13** shows that with varying thickness the tangential strain rates decreases with increases thermal gradient.

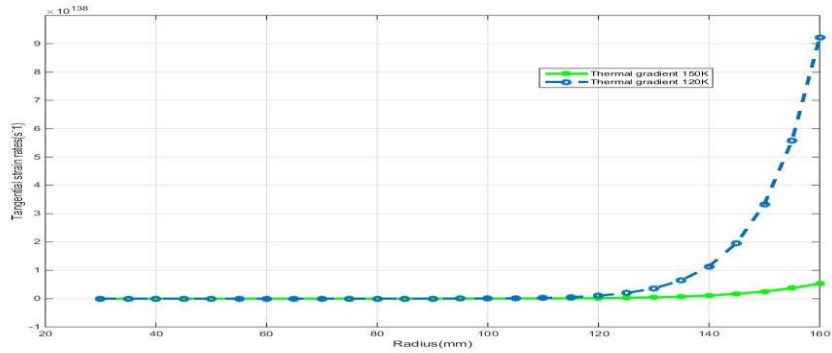

**FIGURE 14**: The variations of the tangential strain rates along radius in rotating composite disk

The graph drawn between the tangential strain rates along radius with thermal gradient 150K and 120K having volume content 10% also with thickness $h_a = 2.50$ and $h_b = 0.50$. The **FIGURE 14** shows that with varying thickness the tangential strain rates decreases with increases thermal gradient. We observed that with increasing volume content the tangential strain rates decreases slightly show in figure14 as compared to **FIGURE13**.

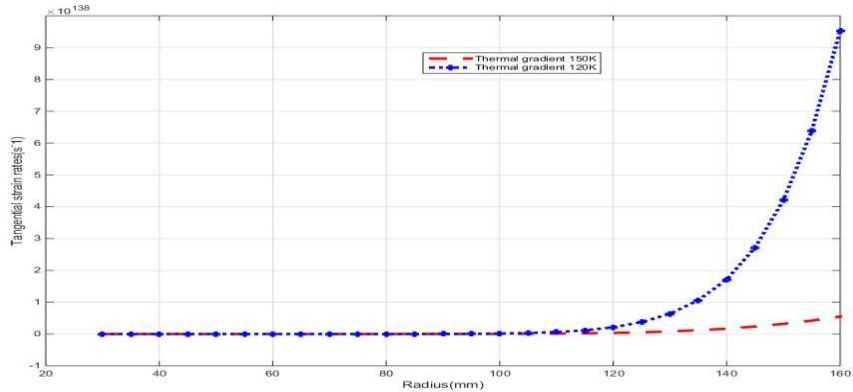

**FIGURE 15**: The variations of the tangential strain rates along radius in rotating composite disk

The graph drawn between the tangential strain rates along radius with thermal gradient 150K and 120K having volume content 5% also with thickness $h_a = 44.22$ and $h_b = 12.97$. The **FIGURE15** shows that with varying thickness the tangential strain rates decreases with increases thermal gradient.

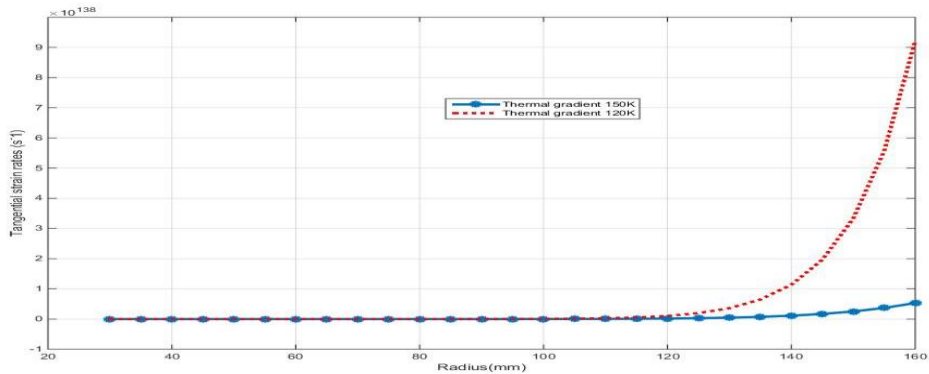

**FIGURE 16**: The variations of the tangential strain rates along radius in rotating composite disk

The graph drawn between the tangential strain rates along radius with thermal gradient 150K and 120K having volume content 10% with thickness $h_a = 44.22$ and $h_b = 12.97$. The **FIGURE16** shows that with varying thickness the tangential strain rates decreases with increases thermal gradient. We observed that with increasing volume content for varying thickness the tangential strain rates decreases slightly shown in **FIGURE16** as compared to **FIGURE15**.

## CONCLUSION

Conclusions based on the graphical representations are:
1. The **FIGURE 1,2,3,4** shows that the radial stresses decreases with increases volume content having varying thickness as well as thermal gradient. We consider thermal gradient 150K and 120K in **FIGURE 1,2** volume content 5% and 10% with thickness $h_a = 2.50$ and $h_b = 0.50$ but in **FIGURE 3,4** volume content 5% and 10% with thickness $h_a = 44.22$ and $h_b = 12.97$ in **FIGURE 3,4** the radial stresses decreases as compare to **FIGURE 1,2**.
2. The **FIGURE 5,6,7,8** shows that the tangential stresses decreases with increases volume content having varying thickness as well as thermal gradient. We consider thermal gradient 150K and 120K in **FIGURE 5,6** volume content 5% and 10% with thickness $h_a = 2.50$ and $h_b = 0.50$ but in **FIGURE 7,8** volume content 5% and 10% with thickness $h_a = 44.22$ and $h_b = 12.97$ in figure 5,6 the stress decreases as compare to **FIGURE 7,8**.
3. The **FIGURE 9,10,11,12** shows that radial strain rates increases with varying thickness $h_a = 2.50$ and $h_b = 0.50$ in **FIGURE 9,10** and $h_a = 44.22$ and $h_b = 12.97$ in **FIGURE 11,12** the tangential stresses slightly in **FIGURE 11,12** as compared to **FIGURE 9,10.**
4. The **FIGURE 13,14,15,16** shows that tangential strain rates decreases with increases thermal gradient varying thickness as well as thermal gradient. We take thickness $h_a = 2.50$ and $h_b = 0.50$ in **FIGURE 13,14** and $h_a = 44.22$ and $h_b = 12.97$ in **FIGURE 15,16** we concluded that the tangential stresses in **FIGURE 15,16** slightly decreases as compared to **FIGURE 13,14**.

It has been concluded that the rotating composite disk with thermal gradient as well as particle gradient having varying thickness help to reduce distortion in the disk.